\documentclass[prd,aps,showpacs,twocolumn,preprintnumbers,amsmath,amssymb,superscriptaddress,nofootinbib, longbibliography]{revtex4-1}

\usepackage{color}
\usepackage{xcolor}
\usepackage{graphicx}
\usepackage[normalem]{ulem}
\usepackage{MnSymbol}
\usepackage{combelow}

\usepackage{hyperref}
\hypersetup{
    colorlinks=true,
    linkcolor=red,
    filecolor=magenta,      
    urlcolor=blue,
    citecolor = teal,
}

\definecolor{limegreen}{RGB}{154,205,0}
\definecolor{brickred}{rgb}{0.8, 0.25, 0.33}
\definecolor{amethyst}{rgb}{0.6, 0.4, 0.8}
\definecolor{azure}{rgb}{0.0, 0.5, 1.0}
\definecolor{awesome}{rgb}{1.0, 0.13, 0.32}
\definecolor{purple}{rgb}{0.8,0,0.6}
\newcommand{\replacea}[2]{{{\color{orange}{#1}}{\color{olive}{\ifmmode\text{\sout{\ensuremath{#2}}}\else\sout{#2}\fi}}}}

\newcommand{\avr}[1]{\left\langle #1 \right\rangle}
\newcommand{\aavr}[1]{\left \llangle #1 \right \rrangle}
\newcommand{\eq}[1]{(\ref{#1})}
\newcommand{\bs}{\boldsymbol}

\begin{document}

\title{Negative Barnett effect, negative moment of inertia of 
gluon plasma and thermal evaporation of chromomagnetic condensate}

\author{Victor~V.~Braguta} 
\email{vvbraguta@theor.jinr.ru}
\affiliation{Bogoliubov Laboratory of Theoretical Physics, Joint Institute for Nuclear Research, 141980 Dubna, Russia}
\affiliation{Moscow Institute of Physics and Technology, 141700 Dolgoprudny, Russia}
\author{Maxim~N.~Chernodub}
\email{maxim.chernodub@univ-tours.fr}
\affiliation{Institut Denis Poisson UMR 7013, Universit\'e de Tours, 37200 Tours, France}
\affiliation{Nordita, Stockholm University, Roslagstullsbacken 23, SE-106 91 Stockholm, Sweden}
\author{Ilya~E.~Kudrov}
\email{ilyakudrov@yandex.ru}
\affiliation{Institute for High Energy Physics, NRC ``Kurchatov Institute'', 142281 Protvino, Russia}
\author{Artem~A.~Roenko}
\email{roenko@theor.jinr.ru}
\affiliation{Bogoliubov Laboratory of Theoretical Physics, Joint Institute for Nuclear Research, 141980 Dubna, Russia}
\author{Dmitrii~A.~Sychev}
\email{sychev.da@phystech.edu}
\affiliation{Bogoliubov Laboratory of Theoretical Physics, Joint Institute for Nuclear Research, 141980 Dubna, Russia}
\affiliation{Moscow Institute of Physics and Technology, 141700 Dolgoprudny, Russia}

\date{\today}

\begin{abstract}
We discuss the negativity of the moment of inertia of (quark-)gluon plasma in a window of ``supervortical'' range of temperatures above the deconfining phase transition, $T \simeq (1\dots 1.5) T_c $ found recently in numerical Monte Carlo simulations by two independent methods. In our work, we confirm numerically that the origin of this effect is rooted in the thermal evaporation of the non-perturbative chromomagnetic condensate. We argue that the negative moment of inertia of gluon plasma indicates the presence of a novel effect, the negative spin-vortical coupling for gluons resulting in a negative gluonic Barnett effect: the spin polarization of gluons exceeds the total angular momentum of rotating plasma, thus forcing the orbital angular momentum to take negative values in the supervortical range of temperatures.
\end{abstract}

\maketitle

%
%
\section{Introduction}

The moment of inertia $I$ of a physical body quantifies the angular momentum $\boldsymbol L$ carried by the body when it is set in rotation with an angular velocity $\bs \Omega$~\cite{Landau_Mechanics_1976}. For slowly rotating rigid mechanical systems, these quantities are related to each other via the linear relation, ${\boldsymbol L} = {\hat I} {\boldsymbol \Omega}$, where the moment of inertia takes the form of the tensor of the second rank, $\hat I$. If the body rotates around one of its principal axes of inertia, the above relation simplifies to: 
\begin{align}
    L = I \Omega,
    \label{eq_L_Omega}
\end{align}
where $I$ is the corresponding eigenvalue of the tensor $\hat I$.

In classical systems residing in thermal equilibrium, the moment of inertia is always a positive quantity, $I > 0$. This intuitively clear statement originates from the fact that a system of physical particles has an everywhere-positive energy density, implying that the momentum of any small subsystem of a rigidly rotating body points out in the direction of its velocity, expressed, in turn, via the angular velocity $\bs \Omega$. As all physical objects we know have a positive energy density, our intuition tells us that they must also have a positive moment of inertia.\footnote{An academic counterexample to this assertion is provided by the Casimir effect~\cite{Casimir:1947kzi}, where the moment of inertia associated with the negative Casimir energy is a negative quantity~\cite{Chernodub:2012ry, Chernodub:2012em} in consistency with the equivalence principle established by the gravitational response of the Casimir energy in the background gravitational field~\cite{Fulling:2007xa, Milton:2007ar}. We do not consider the Casimir effect since it does not play a role in the effect discussed in our article.} 

However, the numerical simulations performed by two different methods, both for static~\cite{Braguta:2023kwl} and rotating~\cite{Braguta:2023yjn} gluon plasmas show that the assertion of the positivity of the mechanical moment of inertia cannot be applied to plasma of gluons. On the contrary, the gluon plasma possesses a negative moment of inertia in a wide window of temperatures above the deconfinement transition~\cite{Braguta:2023kwl, Braguta:2023yjn}. This conclusion drastically contradicts our intuition since (quark-)gluon plasma has a positive energy density at every point while possessing at the same time a negative moment of inertia. According to Ref.~\cite{Braguta:2023yjn}, the origin of the negative moment of inertia lies in the particularities of the evaporation of the chromomagnetic gluon condensate at high temperatures.

A signature of the opposite direction of the angular momentum of the QCD {\it vacuum} with respect to the angular velocity, which would be consistent with $I<0$ in Eq.~\eq{eq_L_Omega}, has also been noticed in non-renormalized data of Ref.~\cite{Yamamoto:2013zwa} about a decade ago. In that work, this property has been attributed to a classical frame-dependence effect as a particle that rests in the laboratory seems oppositely rotating from the point of view of the rotating frame. This observation -- made for the QCD vacuum which is characterized by the gluon and chiral condensates~\cite{Shifman_1978bx, Shifman_1978by} -- lies in the line of our earlier results~\cite{Braguta:2023kwl, Braguta:2023yjn} that highlight the importance of the chromomagnetic condensate for rotation in (quark-)gluon plasma. 

In our work, we study the contribution of the gluon condensate to the moment of inertia of gluon plasma, pointing out the important role of the chromomagnetic component as compared to the ordinary mechanical term. Rotating quark-gluon plasma (QGP) is routinely produced in relativistic heavy-ion collisions~\cite{Abelev2007, STAR_2017ckg}. The appropriate quantity, in this case, is vorticity~\cite{Florkowski:2018fap, Huang:2020dtn, Becattini:2022zvf}, which describes the local circular motion of plasma constituents at a given point of spacetime. 
While simulations show in vortical plasma, vorticity does not generally correspond to a rigid rotation~\cite{Huang:2020dtn}, the rigid rotation is a valuable probe of the response of the quark-gluon plasma to vorticity in most analytical~\cite{Ambrus:2014uqa, Chen:2015hfc, Jiang:2016wvv, Chernodub:2016kxh, Wang:2018sur, Zhang:2020hha, Sadooghi:2021upd, Fujimoto:2021xix, Golubtsova:2021agl, Chen:2022smf, Zhao:2022uxc, Chernodub:2020qah, Mameda:2023sst, Sun:2023dwh, Satapathy:2023oym, Braga:2023qej} and numerical~\cite{Yamamoto:2013zwa, Braguta:2020biu, Braguta:2021jgn, Braguta:2022str, Braguta:2023yjn, Yang:2023vsw} approaches to its thermodynamic properties. Our knowledge of this system can be tested by confronting the analytical results with numerical simulations for a rigid rotation. We follow the same strategy in our paper. 

The paper is organized as follows. In Section~\ref{sec_decomposition}, we discuss the general description of the rotating systems in statistical mechanics and apply it to the rotating gluon matter, paying attention to the existence of the non-perturbative condensates in this strongly interacting system. We show how the gluon moment of inertia can naturally be decomposed into the mechanical and condensate parts. Then, in Section~\ref{sec_results}, we describe our numerical results on the specific moment of inertia.

Section~\ref{ref_condensates} is devoted to the numerical study of a relation between rotation and gluon condensates. In this section, we are also confronting the numerical results of rotation of a strongly interacting, non-perturbative gluon plasma with analytical results on the rotation of the free photon gas. In section~\ref{ref_Barnett}, we present our understanding of the numerical results from the point of view of the distribution of the angular momentum between spin and orbital components of gluons. We suggest that our results point out the existence of the unexpected phenomenon, the negative Barnett effect, which implies that the total angular momentum and orbital angular momentum are pointing in opposite directions due to the strong polarization of the gluon spin. The last section is devoted to the summary of our numerical results and possible implications to the physics of vortical quark-gluon plasma.

%
%
\section{Decomposing moment of inertia of gluons}

\label{sec_decomposition}

\subsection{Moment of inertia in statistical mechanics}

A mechanical reaction of a thermodynamic ensemble to a rigid rotation with the angular velocity $\bs \Omega$ can be characterized by the total angular momentum $\bs J$, which is a thermodynamically conjugated variable with respect to $\bf \Omega$. In systems of particles with spin, such as gluons and quarks, total angular momentum $\bs J$ includes orbital and spin parts. The angular velocity $\bs \Omega$ and the angular momentum $\bs J$ determine the relation between the energy of the rotating system $E^{\text{(lab)}}$ determined in the inertial laboratory frame and the energy $E = E^{\text{(lab)}} - {\bs J} {\bs \Omega}$ corresponding in the non-inertial co-rotating reference frame~\cite{Landau_Mechanics_1976}.

The angular momentum ${\bs J}$ can be expressed via either the energy $E$ (at fixed entropy $S$) or the free energy $F = E - T S$ (at fixed temperature $T$) in the co-rotating frame:
\begin{align}
{\bs J} = - {\left( \frac{\partial E}{\partial {\bs \Omega}} \right)}_{S} = - {\left( \frac{\partial F}{\partial {\bs \Omega}} \right)}_{T}\,.
\label{eq_L_via_tildeF}
\end{align}
Here, we used the thermodynamic relations 
\begin{align}
    d E = T d S - {\bs J} d {\bs \Omega}\,, \qquad d F = - S d T - {\bs J} d {\bs \Omega}\,,
\end{align}
to arrive at Eqs.~\eqref{eq_L_via_tildeF}.

The moment of inertia $I \equiv I_{\bs n}$ of a physical system with respect to global rotations with the angular velocity ${\bs \Omega} = \Omega {\bs n}$ around the principal axis of inertia
$\bs n$ (with ${\bs n}^2 = 1$) is determined via its free energy~$F$ in the co-rotating reference frame as follows~\cite{Landau_Statistics_1976}:
\begin{align}
    I(T) = \frac{J(T,\Omega)}{\Omega} {\biggl|}_{\Omega \to 0} \equiv - \frac{1}{\Omega} {\left( \frac{\partial F(T,\Omega)}{\partial \Omega} \right)}_{T}{\biggl|}_{\Omega \to 0}\,.
\label{eq_I_F}
\end{align}

Since we are working with a relativistic system, it is important to stress that we implicitly imply that the rigid rotation does not violate the causality. In practice, this requirement implies that the size of the body in the direction perpendicular to the axis should be bounded, $R_\perp < R_{\mathrm{light}}$, by the radius of the light cylinder, $R_{\mathrm{light}} = 1/|\Omega|$. Equation~\eq{eq_I_F} also allows us to work in the thermodynamic limit $R_\perp \to \infty$, provided it is taken with a proper order of limits, $\lim_{R_\perp \to \infty}\lim_{\Omega \to 0}$. Alternatively, one can take the radius of the system $R_{\perp} = C_0/\Omega$ with the real-valued parameter $0 < C_0 < 1$ to support the causality property in the system.

Our experience in classical mechanics tells us that we do not need to set a physical system in motion for the purpose of evaluating its moment of inertia. In other words, the moment of inertia can also be calculated in the static limit when the object does not rotate at all. This observation simplifies the calculation of the moment of inertia since it does not require introduction of the curved space with a nontrivial metric (the latter method has been actively used, for example, in Refs.~\cite{Yamamoto:2013zwa, Braguta:2020biu, Braguta:2021jgn, Braguta:2022str, Braguta:2023yjn, Yang:2023vsw}). In our paper, we analyze the moment of inertia of (quark-)gluon plasma using standard, nonrotating lattices for which the model can conveniently be formulated in a flat Euclidean spacetime following the strategy of Ref.~\cite{Braguta:2023kwl}.

\subsection{Moment of inertia of gluons and chromomagnetic condensate}

The moment of inertia~\eq{eq_I_F} of the gluon plasma can be decomposed into two parts~\cite{Braguta:2023yjn}:
\begin{align}
    I^{\mathrm{gl}} = I^{\mathrm{gl}}_{\text{mech}} + I^{\mathrm{gl}}_{\text{magn}}\,.
    \label{eq_I}
\end{align}
This result can be readily obtained from Eq.~\eq{eq_I_F} using the free energy 
\begin{align}
F = - T \ln \int D A e^{i S}\,,
\label{eq_F_Minkowski}
\end{align}
of the rotating gluon gas described by Yang-Mills theory with the action in Minkowski spacetime: 
\begin{align}
S = - \frac{1}{2 g^{2}_{\rm YM}} \int d^{4} x\, \sqrt{- g}\,  g^{\mu \nu} g^{\alpha \beta} F_{\mu \alpha}^{a} F_{\nu \beta}^{a}\,.
\label{eq_S_Minkowski}
\end{align}
The rotation (taken to be around the $z$-axis) is defined by the following metric tensor:
\begin{align}
g_{\mu \nu} = 
\begin{pmatrix}
1 - r_\perp^2 \Omega^2 & \Omega y & -\Omega x & 0 \\
\Omega y & -1 & 0 & 0  \\ 
-\Omega x & 0 & -1 & 0 \\
0 & 0 & 0 & -1
\end{pmatrix}\,,
\label{eq_metric}
\end{align}
with $r_\perp = \sqrt{x^2 + y^2}$.
Action~\eqref{eq_S_Minkowski} in the curved space with metric~\eqref{eq_metric} is a parabolic function of the angular frequency~$\Omega$:
\begin{equation}
S = S_0 + S_1 \Omega + \frac{S_2}{2} \Omega^2\,,
\label{eq_S_Minkowski_structure}
\end{equation}
where $S_0$ is the action of gluons without rotation and
\begin{align}
    S_1 = {}& \frac {1} {g_{Y M}^2} \int d^4 x \bigl [
    x F^a_{yx} F^a_{xt} +xF^a_{yz} F^a_{zt} - {} \nonumber \\
    &\hspace{7em} {} -  y F^a_{xy} F^a_{yt} -y F^a_{xz} F^a_{zt}
    \bigr ], 
    \label{eq:s1} \\
     S_2 = {}& -\frac 1 {g_{Y M}^2} \int d^4 x \bigl [
    r_\perp^2 (F^a_{xy})^2 + y^2 (F^a_{xz})^2 + {} \nonumber \\
    &\hspace{6em} {} + x^2 (F^a_{yz})^2 + 2xy F^a_{xz} F^a_{zy}
     \bigr ]. \label{eq:s2}
\end{align}
To ensure the decomposition~\eqref{eq_I} one can
substitute the gluon action~\eqref{eq_S_Minkowski_structure} into the definition of the free energy~\eqref{eq_F_Minkowski} and perform the double differentiation with respect to the angular frequency $\Omega$ in order to obtain the moment of inertia $I$ as described in the second relation of Eq.~\eqref{eq_I_F}. Since action~\eqref{eq_S_Minkowski_structure} includes both linear and quadratic parts in $\Omega$, one can expect the appearance of a two-point correlator term and a one-point local term in the static limit, $\Omega \to 0$. According to Eq.~\eqref{eq_I_F}, the limit should be implemented after the differentiation over the angular frequency. 
The mentioned terms are, respectively, the mechanical and magnetic contributions shown in Eq.~\eqref{eq_I}. Now, let us consider their physical meaning in detail.

The first term in the right-hand side of Eq.~\eq{eq_I}:
\begin{align}
    I^{\mathrm{gl}}_{\text{mech}} = T \aavr{S_1^2}_T = \frac{1}{T} \aavr{ \bigl({\bs n} \cdot {\bs J}^{\mathrm{gl}} \bigr)^2}_T, \quad
\label{eq_I_mech}
\end{align}
is the mechanical contribution which accounts for the fluctuations of the total angular momentum of gluons,
\begin{align}
        J_i^{\mathrm{gl}} = \frac{T}{2} \int d^4 x \, \epsilon_{ijk} M^{jk}_{\mathrm{gl}}(x)\,,
        \qquad i,j = 1,2,3\,,
        \label{eq_angular_momentum}
\end{align}
projected on the axis of rotation $\bs n$. The local angular momentum of gluons,
\begin{align}
    M^{ij}_{\mathrm{gl}}({ x}) = x^i T^{j0}_{\mathrm{gl}}({ x}) - x^j T^{i0}_{\mathrm{gl}}({ x})\,,
    \label{gl_ang_mom}
\end{align}
is expressed through the gluonic stress-energy tensor:  $T^{\mu\nu}_{\mathrm{gl}} = F^{a,\mu\alpha} F^{a,\nu}_{\quad\, \alpha} - (1/4) \eta^{\mu\nu} F^{a,\alpha\beta} F^a_{\alpha\beta}$, with the metric $\eta_{\mu\nu} = {\text{diag}}(+1,-1,-1,-1)$ of flat Minkowski spacetime. Here, we use notations 
\begin{align}
    \aavr{{\mathcal O}}_T = \left\langle {\mathcal O} \right\rangle_T - \left\langle {\mathcal O} \right\rangle_{T=0}\,,
    \label{eq_subtraction}
\end{align}
to represent the thermal part of the expectation value of an operator $\mathcal O$. We also notice that $\langle {\bs n} \cdot {\bs J} \rangle \equiv 0$ at ${\bs \Omega} = 0$. 

In the thermodynamic sense, the moment of inertia of a system is associated with its susceptibility with respect to global rotations. For example, Eq.~\eq{eq_I_mech} has a form of a susceptibility, which estimates how the system is susceptible to rotational motion. 

Thermodynamic susceptibilities are associated with the second-order derivatives of the free energy with respect to the corresponding conjugate variables (for example, with respect to a chemical potential if we study density fluctuations). Since rotation is defined by the shift of the Hamiltonian, $H_0 \to H = H_0 - {\bs \Omega} \cdot {\bs J}$, the second-order derivative of the corresponding free energy\footnote{Equation~\eq{eq_I_F} can be rewritten as a second derivative since $\lim\nolimits_{x \to 0} f'(x)/x =  f''(0)$ for any function $f(x)$, continuous in the origin and possessing a vanishing first derivative, $f'(0) = 0$.} 
$F = - T \ln {\mathrm{Tr}}\, e^{- H/T}$ with respect to the angular frequency $\Omega$, gives us susceptibility~\eq{eq_I_mech} of the angular momentum ${\bs J}$ provided the latter does not depend on the angular velocity ${\bs \Omega}$. For the gluon field considered in our article, the angular momentum depends explicitly on the angular velocity ${\bs \Omega}$. This property leads to the second term in Eq.~\eq{eq_I}.

The second contribution to the total momentum~\eq{eq_I} is given by the fluctuation of the chromomagnetic field~\cite{Braguta:2023yjn}:
\begin{align}
 I^{\mathrm{gl}}_{\text{magn}} & {} = T \aavr{S_2}_T = \nonumber \\   = & \int_V d^3 x \Bigl[\aavr{({\bs B}^a \cdot {\bs x}_\perp)^2}_T + \aavr{({\bs B}^a \cdot {\bs n})^2}_T {\bs x}_\perp^2 \Bigr]\,,  
\label{eq_I_magn}
\end{align}
where $B^a_i = \frac{1}{2} \epsilon^{ijk} F_{jk}^a$ is the chromomagnetic field and ${\bs x}_\perp = {\bs x} - {\bs n} ({\bs n} \cdot {\bs x})$ is the vector between the point $\bs x$ and the axis of rotation set by the unit vector ${\bs n}^2 = 1$. The normalization of Eq.~\eq{eq_I} is chosen in such a way that cold ($T = 0$) vacuum has no moment of inertia with both mechanical~\eq{eq_I_mech} and magnetic~\eq{eq_I_magn} contributions vanishing.\footnote{Therefore, a proper renormalization of the data of Ref.~\cite{Yamamoto:2013zwa} would give us a zero moment of inertia for the QCD vacuum at zero temperature as nothing (i.e., vacuum) cannot move or rotate.} 

While the mechanical correlator term~\eq{eq_I_mech} always gives a positive contribution to the moment of inertia, the local magnetic term~\eq{eq_I_magn} is proportional to the magnetic gluonic condensate, which does change the sign above the deconfinement phase transition~\cite{Boyd:1996bx}. It is the last term that is responsible for the negative moment of inertia of the gluon plasma~\cite{Braguta:2023kwl, Braguta:2023yjn}. 

Both terms in Eq.~\eq{eq_I} are calculated in the static limit, ${\bs \Omega} \equiv 0$, which possesses the $\mathrm{SO}(3)$ rotational symmetry. For the magnetic term~\eq{eq_I_magn}, the space isotropy implies
\begin{align}
\aavr{B^a_i B^a_j}_T = \frac{1}{3} \delta_{ij} \aavr{({\bs B}^a)^2}_T\,,
\label{eq_SO3_symmetry}
\end{align}
thus allowing us to link the magnetic contribution~\eq{eq_I_magn} to the magnetic component of the gluon condensate:
\begin{align} \hskip -1mm
    I^{\mathrm{gl}}_{\text{magn}} = \frac{2}{3}\int_V d^3 x \, {\bs x}_\perp^2 \aavr{({\bs B}^a)^2}_T 
\label{eq_I_magn_2}
\end{align}
Surprisingly, the chromomagnetic contribution~\eq{eq_I_magn_2} has the same form as the moment of inertia of a classical body with the mass density $\rho({\bs x})$, rotating about one of its principal inertial axes. Indeed, Eq.~\eq{eq_I_magn_2} can be written in the form of an integral over the volume $V$ of the system:
\begin{align}
    I_{\mathrm{class}} = \int_V d^3 x \, {\bs x}_\perp^2 \, \rho({\bs x}) \,.
    \label{eq_classical}
\end{align}
One recovers the non-perturbative quantum contribution of the gluon condensate~\eq{eq_I_magn_2} from the classical expression~\eq{eq_classical}, with the help of the formal substitution: $\rho({\bs x}) \to \frac{2}{3} \aavr{({\bs B}^a)^2}_T$. While this relation might seem natural, we will see below that $\aavr{({\bs B}^a)^2}_T < 0$ in a substantial range of temperatures above $T_c$, which, in the classical formula~\eq{eq_classical}, would correspond to an unphysical negative mass density, $\rho({\bs x}) < 0$. Of course, it is well known that the gluon plasma does not have a negative mass (energy) density~\cite{Boyd:1996bx} and still, we show~\cite{Braguta:2023kwl, Braguta:2023yjn} the moment of inertia of this plasma becomes a negative quantity in a range of temperatures. We will discuss this would-be paradox below.

%
%
\section{Specific moment of inertia: results of numerical simulations} 
\label{sec_results}

\subsection{Thermodynamics and continuum limit}

Before discussing the results of our numerical simulations, we would like to comment on their reliability in describing physics in the continuum limit. 

Given the extensivity of the moment of inertia $I$ and, consequently, its dependence on the shape of the system, it is convenient to consider an associated non-extensive quantity $K_2 = - {I}/{F_0 R_\perp^2}$, which represents a dimensionless moment of inertia. This expression, written for a cylinder-shaped object, is normalized by its size (squared) in transverse dimensions $R_\perp$ as well as by the free energy in the nonrotating limit, 
\begin{align}
    F_0 = \lim_{\Omega \to 0} F < 0\,. 
\end{align}
For a non-interacting system of massless particles, one gets the exact result: $K_2 = 2$~\cite{Ambrus:2023bid}. This result is also recovered, within the numerical accuracy, in the high-temperature limit of quark-gluon plasma~\cite{Braguta:2023yjn}, which shows the consistency of our understanding of the numerical data. 

The independence of the dimensionless moment of inertia $K_2$ on the spatial volume of the system (with various spatial sizes of the lattice, $N_s$), established in Ref.~\cite{Braguta:2023yjn}, implies that our results are very close to the thermodynamic limit. In addition, the excellent scaling towards the continuum limit, taken as a sequence of various $N_t$, has been established~\cite{Braguta:2023kwl, Braguta:2023yjn}. Given these observations, we choose below to work only with a single lattice spacing (corresponding to the temporal extension $N_t = 6$) and tree-level improved Symanzik lattice gauge action, which is adequately close to continuum physics.

The moment of inertia is an extensive quantity that behaves as $I \propto V R_\perp^2$, where $V$ is the volume of the system and $R_\perp$ is the size of the system with respect to the axis of rotation. Since we work on square lattices with the transverse extension $R_\perp = L_s/2$, we will present the results of the specific moment of inertia defined as
\begin{align}
    {\mathsf i}_2 = \frac{I}{V R_\perp^2}\,, \qquad R_\perp = \frac{L_s}{2}\,.
    \label{eq_i2}
\end{align}
Here the specific (size-independent) notations are introduced as ${\mathsf i}_n = I_n/(V R_\perp^n)$, with $n=2,4, \dots$, in order to highlight the size-dependence of the expansion of the free energy $F$, defined in the co-rotating reference frame, over the angular frequencies $\Omega$:
\begin{align}
        F(T, R_\perp, \Omega) & = F_0(T, R_\perp) - V \sum_{k=1}^\infty \frac{{\mathsf i}_{2k}(T)}{(2k)!}  R_{\perp}^{2k} \Omega^{2k} \nonumber \\
        & \equiv F_0 - \frac{I}{2} \Omega^2 + O(\Omega^4) \,,
    \label{eq_series}
\end{align}
(see, e.g., Ref.~\cite{Ambrus:2023bid}). The usual moment of inertia, $I_2 \equiv I$, corresponding to the static limit $\Omega \to 0$, appears in the second term in the series in the second line of Eq.~\eq{eq_series}. It is expressed via the specific moment of inertia ${\mathsf i}_2$ defined in Eq.~\eq{eq_i2}. The higher-order terms in Eq.~\eq{eq_series} correspond to the nonlinear contributions to the moment of inertia due to the deformation of the system caused by rotation. We will not consider them in this article. For a rotating system of free bosons, these terms were computed analytically in Ref.~\cite{Ambrus:2023bid}.

\subsection{Setup of numerical simulations}

Both mechanical and chromomagnetic contributions to the total moment of inertia~\eqref{eq_I} may be calculated separately in lattice simulations. Following the numerical strategy of Ref.~\cite{Braguta:2023kwl}, 
our calculation is based on 
representation~\eqref{eq_I_mech},~\eqref{eq_I_magn} of the contributions to specific moment of inertia. The discretized lattice expressions for operators $S_1$, $S_2$, shown in the continuum theory in Eqs.~\eqref{eq:s1} and~\eqref{eq:s2}, can be found in Ref.~\cite{Braguta:2021jgn}.
In the lattice framework, a proper zero-temperature subtraction in Eqs.~\eqref{eq_I_mech},~\eqref{eq_I_magn} is performed for specific quantities. We calculate these two parts of the moment of inertia on the finite temperature nonrotating lattice $6\times 24\times 31^2$ with the zero-temperature subtraction performed at $N_t = 24$. We use the tree-level improved Symanzik lattice gauge action, while other details of the lattice simulations may be found in Ref.~\cite{Braguta:2023kwl}. Note that in this lattice setup, the specific moment of inertia does not contain any divergences and supports the continuum limit extrapolation with the finite result~\cite{Braguta:2023kwl}.

\subsection{Mechanical and magnetic contributions}

\begin{figure}[t!]
    \includegraphics[width = 0.98\linewidth]{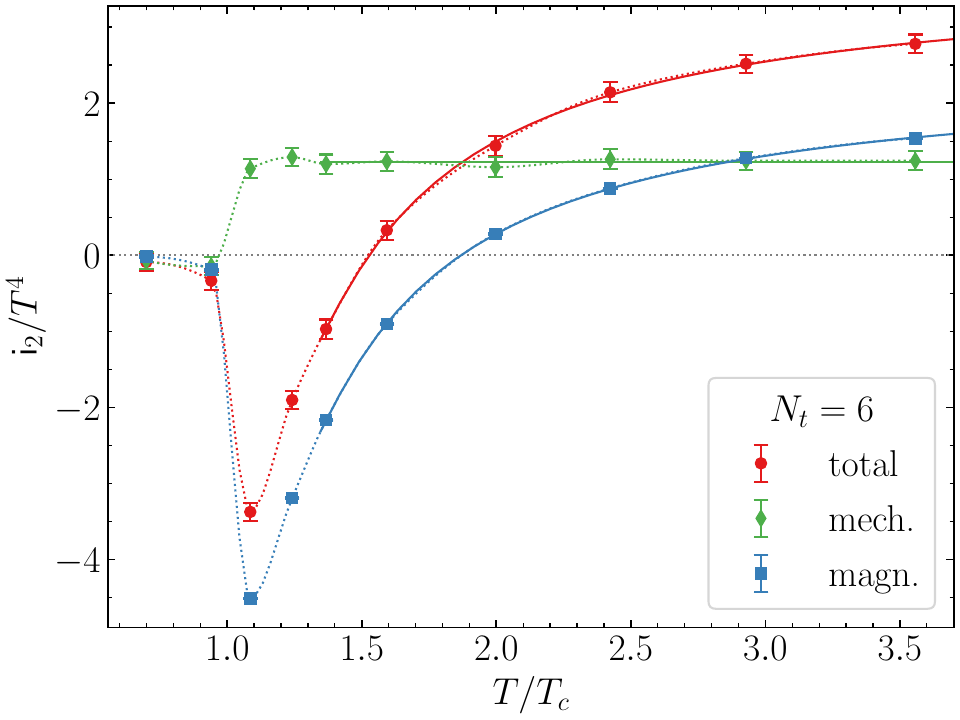}
    \caption{The mechanical~\eq{eq_I_mech} and chromomagnetic~\eq{eq_I_magn} contribution to the total moment of inertia~\eq{eq_I_specific} along with the fits~\eq{eq_mech_fit} and \eq{eq_magn_fit}, respectively (given by the solid lines). The dashed lines are plotted to guide the eye.}
    \label{fig_I}
\end{figure}

Monte Carlo results for the specific moment of inertia of SU(3) gluon plasma are presented in Fig.~\ref{fig_I}. We plot separately the mechanical term~\eq{eq_I_mech}, the magnetic term~\eq{eq_I_magn}, and the total {\it specific} moment of inertia~\eq{eq_I}:
\begin{align}
    {\mathsf i}_2 = {\mathsf i}_2^{\mathsf{mech}} + {\mathsf i}_2^{\mathsf{magn}}.
    \label{eq_I_specific}
\end{align}
How should we understand these results?

%
%

Deeply in the confinement (hadronic) phase, the moment of inertia of the system is zero. This property has a simple explanation since there is no physical matter that can be set in motion at a temperature substantially lower than $T_c$. Indeed, intuitively, we cannot move or rotate a true ``nothing'' (i.e., vacuum). We have no fluid or gas matter to stir, so solid crystal to torque into rotation (see also discussion in Refs.~\cite{Chen:2015hfc, Chernodub:2016kxh}). The vanishing of the moment of inertia is well seen in Fig.~\ref{fig_I} for the temperature point $T \simeq 0.70 T_c$. This result also demonstrates the correctness of the additive normalization procedure~\eq{eq_subtraction} used in our calculations.

As soon as the temperature rises and we are approaching the deconfining temperature, glueballs become thermally activated, and they form an interacting dilute gas. In addition, since the phase transition in SU(3) Yang-Mills theory is of a (weak) first order, the thermal fluctuations should produce small bubbles of the high-temperature phase. Thus, slightly below the critical temperature $T_c$, we have a warm gluon matter, which is expected to be sensitive to rotation and possess a small moment of inertia. Our results in Fig.~\ref{fig_I} at $T \simeq 0.94 T_c$ confirm this anticipated fact, which, however, possesses one surprising property that the total moment of inertia has a tendency to become a negative quantity in agreement with~\cite{Braguta:2023kwl, Braguta:2023yjn}. 

Figure~\ref{fig_I} shows that this extraordinary property extends also to the gluon plasma regime above the critical temperature. The total moment of inertia $I$ is negative up to the supervortical critical temperature~\cite{Braguta:2023kwl, Braguta:2023yjn},
\begin{align}
T_s = 1.50(10)T_c\,,
\label{eq_T_s}
\end{align}
at which it vanishes and turns to positive values at $T > T_s$. The positivity of $I$ in the high-temperature regime fits well with our understanding that plasma becomes overwhelmingly perturbative at $T \to \infty$.

The mechanical contribution~\eq{eq_I_mech} possesses a mundane constant behavior with $I_{\mathrm{mech}}^{\mathrm{gl}} > 0$, which gets established right above the critical temperature. The chromomagnetic part~\eq{eq_I_magn} of the moment of inertia is, however, a nontrivial function of temperature. It takes a negative value in a range of temperatures, which makes the total momentum of inertia negative as well in the supervortical region of temperatures $T_c \lesssim T < T_s$.

We work in the formalism of Ref.~\cite{Braguta:2023kwl}, which does not require the introduction of the metric tensor, thus giving more robustness to our calculations. Another way to calculate the moment of inertia on the lattice involves the simulation of the rotating Yang-Mills theory in curved space and an analytic continuation procedure similar to Refs.~\cite{Braguta:2020biu, Braguta:2021jgn, Braguta:2022str, Yang:2023vsw}. The results of such calculations, presented in Ref.~\cite{Braguta:2023yjn}, firmly overlap with them from nonrotating lattices, thus demonstrating the consistency of our approaches.

Now let us address the question of how it can be that the moment of inertia of a gluon gas is negative while the energy density of the very same gluon gas is positive everywhere. The last statement contradicts our experience of classical mechanics, Eq.~\eq{eq_classical}, which implies that the moment of inertia should both be a positive quantity as the mass density is a positive quantity as well.

%
%
\section{Rotation and condensates}
\label{ref_condensates}

\subsection{Thermal contributions to gluon condensates}

The negative moment of inertia originates from the thermal properties of (chromo)magnetic condensate~\eq{eq_I_magn_2}. The chromomagnetic, $\aavr{{\mathcal B}^2}_T$, and chromoelectric, $\aavr{{\mathcal E}^2}_T$, condensates enter the (thermal part of) the total gluon condensate:
\begin{align}
    \aavr{G^2}_T = \aavr{{\mathcal B}^2}_T + \aavr{{\mathcal E}^2}_T\,,
    \label{eq_G2}
\end{align}
which determines, via the trace anomaly~\cite{Boyd:1996bx},
\begin{align}
    \varepsilon - 3 p = \avr{T^\mu_{\ \mu}} = \frac{\beta(\alpha_s)}{4 \pi} \aavr{(F^a_{\mu\nu})^2}_T \equiv - \aavr{G^2}_T\,.
    \label{eq_anomaly}
\end{align}
the equation of state, which relates the energy density $\varepsilon$ with pressure $p$. Eq.~\eq{eq_anomaly} involves the beta function, 
\begin{align}
    \beta\bigl(\alpha_s(\mu)\bigr) = \frac{{\mathrm d} \alpha_s(\mu)}{{\mathrm d} \ln \mu} < 0 \,,
    \label{eq_beta_s}
\end{align}
which reflects the running of the QCD coupling constant $\alpha_s = g^2/(4 \pi)$ with the energy scale $\mu$. 

According to the last relation in Eq.~\eqref{eq_anomaly}, the chromomagnetic part of the gluon condensate~\eqref{eq_G2} --- that enters, for example, the definition of the magnetic contribution to the moment of inertia~\eqref{eq_I_magn_2} --- is related to the expectation value of the chromomagnetic field,
\begin{align}
    \aavr{{\mathcal B}^2}_T = - \frac{\beta(\alpha_s)}{4 \pi} \llangle{({\boldsymbol{B}})}^2\rrangle_T\,,
\end{align}
via the beta function~\eqref{eq_beta_s}. The same relation holds for its chromoelectric counterpart. 

\begin{figure}[!t]
    \includegraphics[width = 0.98\linewidth]{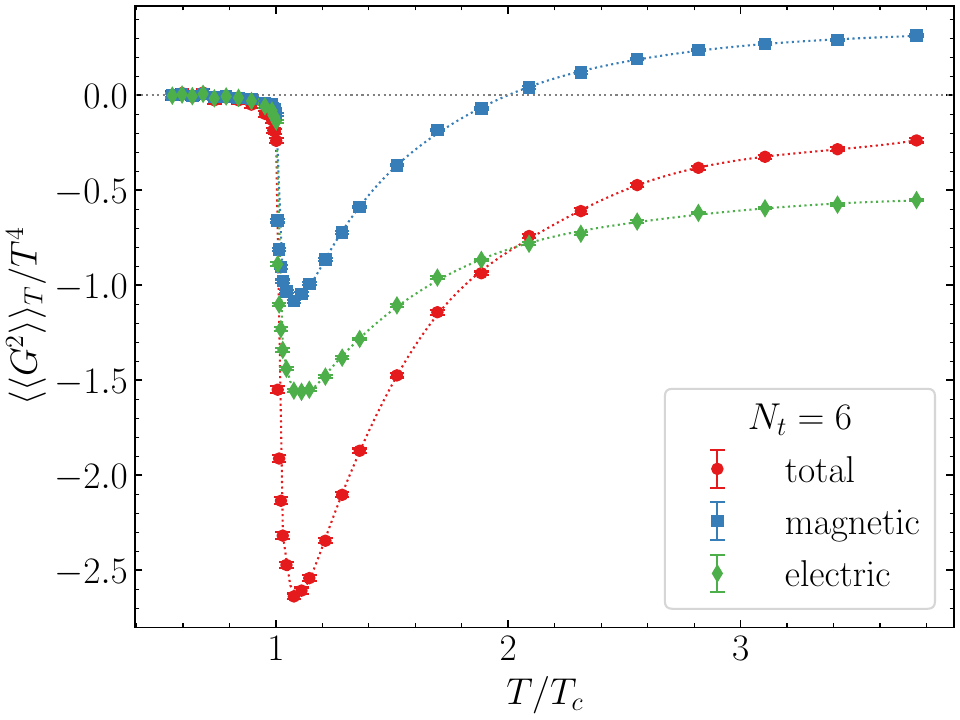}
    \caption{The total gluon condensate~\eq{eq_anomaly}, and, separately, its electric $\aavr{{\mathcal E}^2}_T$ and magnetic $\aavr{{\mathcal B}^2}_T$ components~\eq{eq_G2}, as functions of temperature~$T$.}
    \label{fig_condensates}
\end{figure}

Our Monte Carlo results for all gluonic condensates~\eq{eq_anomaly} are shown in Fig.~\ref{fig_condensates}.\footnote{This figure essentially reproduces Fig.~10 (taken with a minus sign) of Ref.~\cite{Boyd:1996bx} re-calculated for the improved Symanzik action.} These results were calculated within standard lattice technique~\cite{Boyd:1996bx, Borsanyi:2012ve} on the lattice $6\times 36\times 145^2$ with performed zero-temperature subtraction at $N_t = 36$.
The magnetic condensate becomes negative above the phase transition, and then it changes the sign to a positive domain at $T \simeq 2 T_c$~\cite{Boyd:1996bx}.

How can we understand the thermal behavior of the magnetic condensate? One can associate its behavior with the evaporation of the magnetic component of the gluon plasma~\cite{Chernodub_2006gu, Liao_2008jg} and properties of string excitations in the deconfinement phase~\cite{Chernodub_2006gu, Glozman_2022zpy}. Moreover, the magnetic condensate is an infrared non-perturbative property of the vacuum. As with any other condensate, its emergence can be associated with the infrared coherence of the vacuum state. Strong thermal fluctuations add considerable ultraviolet noise to the system and, consequently, disorder this condensate. Therefore, the condensate quickly evaporates as the temperature rises above the critical point, $T = T_c$. This behavior is clearly seen in Fig.~\ref{fig_condensates} in the temperature range $1 \lesssim T/T_c  \lesssim 1.1$. Notice that the magnetic condensate is not an order parameter of the deconfinement phase transition, and, therefore, nothing special happens with it at the phase transition $T = T_c$ apart from the quick drop caused by the evaporation. These features are pertinent both to magnetic and electric parts of the condensate, which affect scale (trace) anomaly and determine, as a consequence, the non-perturbative behavior of the equation of state of the gluon plasma~\cite{Boyd:1996bx} in the deconfinement phase transition. The observed effects are non-perturbative facets of thermal Yang-Mills theory that evade a rigorous treatment so far.

The finite temperature environment also generates essentially perturbative thermal fluctuations in the system, which contribute positively to the magnetic condensate. As temperature increases above $T \simeq 1.1 T_c$, the rise of the magnetic thermal fluctuations overcome the evaporation of the remaining magnetic condensate, and the slope changes: at this temperature, the quantity $\aavr{{\mathcal B}^2}_T$ starts to grow.\footnote{At $T \gtrsim 1.1 T_c$, the quantity $\aavr{{\mathcal B}^2}_T$ includes an essential contribution coming from thermal fluctuations so that it is not justified, strictly speaking, to call it a condensate. We will, however, continue to use the word ``condensate'' to keep our narration smooth.} The thermal contribution to the electric condensate $\aavr{{\mathcal E}^2}_T$ also follows the same pattern, Fig.~\ref{fig_condensates}.

At $T \simeq 2T_c$, the thermal contribution becomes larger than the drop due to the evaporation of the $T=0$ condensate, and $\aavr{{\mathcal B}^2}_T$ becomes a positive quantity at higher temperatures, approaching the expected $\aavr{{\mathcal B}^2}_T \propto T^4$ Stefan-Boltzmann law at $T \to \infty$.

Since the right-hand side of the anomaly equation~\eq{eq_anomaly}, given by the {\it sum} of electric and magnetic condensates, is always a positive quantity; therefore, thermodynamic relations imply that the energy density is always a positive quantity as well~\cite{Boyd:1996bx}. Thus, the moment of inertia, substantially altered by the chromomagnetic component~\eq{eq_I_magn} can take negative values. At the same time, the energy density, given by the sum of both electric and magnetic contributions~\eq{eq_G2}, is always a positive quantity. In other words, one can have a negative moment of inertia for an object with an everywhere-positive energy density.

%
%

\subsection{Confronting rotation of a non-perturbative gluon plasma with free photon gas}

It is instructive to compare our numerical results for gluonic plasma, extrapolated to the high-temperature limit, with thermodynamics of non-interacting massless vector fields, that is, photons. The interest in this comparison originates from the fact that the generic relation~\eq{eq_I_F} equally applies to rotating photon gas. In contrast, the vector nature of the photon field assures -- similarly to the gluon field -- that the photon gas receives two contributions to its moment of inertia~\eq{eq_I}, both mechanical~\eq{eq_I_mech} and magnetic~\eq{eq_I_magn} ones. 

One could naively expect that at high temperatures, the physics of gluons becomes perturbative due to asymptotic freedom: in a hot gluon plasma, the mean energy of gluons is high, and for such high-energy gluons, the Yang-Mills coupling constant is asymptotically vanishing. However, this expectation does not work for all gluons since the magnetic gluon component always contains a non-perturbative part characterized by an effective magnetic, dimensionally reduced coupling, $g^2_{3d} = g^2_{\rm YM}(T) T$, regardless of how high the temperature of the gluon plasma is~\cite{Karsch:2001cy}. 

Moreover, in the studied range of temperatures, the equation of state of the gluon plasma contains a strong correction from a non-perturbative sector~\cite{Karsch:2001cy} and therefore, any relation with free photodynamics is at least far-fetched. In addition, it is the chromomagnetic contribution~\eq{eq_I_magn} that makes the negative contribution to the moment of inertia. However, contrary to the gluon case, the magnetic part of photons is not non-perturbative and easily calculable analytically, which makes the comparison of high-temperature gluodynamics with high-temperature photodynamics especially valuable. Below, we will make the comparison, normalized by degrees of freedom, to see how different the thermodynamics of these rotating vector fields are.

\subsubsection{Photons}

To evaluate the total moment of inertia of rotating photons, one can take the result for a free scalar field~\cite{Ambrus:2023bid} and rescale it by the factor of $2$ corresponding to degrees of freedom of a photon. One gets for a photon in a cylinder of the radius~$R$ and the height $L$:
\begin{align}
    I^{\mathrm{ph}}_{\mathrm{total}} \equiv I^{\mathrm{ph}}_{\mathrm{mech}} + I^{\mathrm{ph}}_{\mathrm{magn}} = \frac{2\pi^2}{45} L R^4 T^4\,.
    \label{eq_I_ph_full}
\end{align}
This expression excellently agrees with the numerical results of Ref.~\cite{Chernodub:2018era} for the moment of inertia of hot photon gas rotating in a cylinder, $I_{\mathrm{ph}}/(L R^4 T^4) \simeq 0.44$.

Similarly to gluons~\eq{eq_I}, the total moment of inertia~\eq{eq_I_ph_full} of the photon gas possesses two contributions which are easily calculable. The mechanical part, given by the susceptibility of the angular momentum~\eq{eq_I_mech},
\begin{align}
    I^{\mathrm{ph}}_{\mathrm{mech}} = 
    \frac{\pi^2}{90} (4 - \pi) L R^4 T^4\,,
    \label{eq_I_ph_corr}
\end{align}
can be obtained by subtraction of the contribution from thermal fluctuations of magnetic field~\eq{eq_I_magn},
\begin{align}
    I^{\mathrm{ph}}_{\mathrm{magn}} \equiv \avr{{\bs B}^2}_T \cdot \frac{2}{3} \int_V d^3 r \, r_\perp^2 = \frac{\pi^3}{90} L R^4 T^4\,,
    \label{eq_I_ph_magn}
\end{align}
from the full expression~\eq{eq_I_ph_full}. In order to calculate Eq.~\eq{eq_I_ph_magn}, we took into account the Stefan-Boltzmann value of the thermal fluctuations of the magnetic field, 
\begin{align}
    \avr{{\bs B}^2}_T = \frac{\pi^2}{15} T^4 \qquad\ {\text{[photon gas]}}\,.
\end{align}
The mechanical contribution~\eq{eq_I_ph_corr}, obtained by the difference between the total value~\eq{eq_I_ph_full} and the magnetic contribution~\eq{eq_I_ph_magn}, gives only a small fraction of the total moment of inertia of the photon gas:
\begin{align}
    \frac{I^{\mathrm{ph}}_{\mathrm{mech}}}{I^{\mathrm{ph}}_{\mathrm{total}}} = 1 - \frac{\pi}{4} \simeq 0.215, \quad {\mathrm{or}} \quad \frac{I^{\mathrm{ph}}_{\mathrm{magn}}}{I^{\mathrm{ph}}_{\mathrm{mech}}} = \frac{\pi}{4 - \pi} \simeq 3.66\,.
\end{align}
This observation should be compared with the behavior of gluon plasma shown in Fig.~\ref{fig_I}, where the magnetic contribution is indeed bigger than the mechanical part, which, however, still takes an essential part of the total moment of inertia at our higher temperature $T \simeq 3.6 T_c$.

\subsubsection{Gluons}

The asymptotic values of the mechanical and magnetic parts of the moment of inertia in the gluon plasma can be obtained by suitable fits. The mechanical part is almost insensitive to temperature in the deconfinement phase, and it can be well described in the region $T \gtrsim 1.35T_c$, with $\chi^2/d.o.f. \simeq 0.1$, by the constant function, 
\begin{align}
\frac{I_{\mathrm{mech}}^{\mathrm{gl}}(T)}{T^4} = \frac{I_{\mathrm{mech}}^{\mathrm{gl},\infty}}{T^4} = 1.22(6)\,. 
\label{eq_mech_fit}
\end{align}
The chromomagnetic contribution, as well as the total specific moment of inertia, has a more involved behavior which can be fitted at $T \gtrsim 1.35 T_c$ by the following function (with $\chi^2/{\mathrm{d.o.f.}}  \simeq 1.9$ for the magnetic part and $\chi^2/{\mathrm{d.o.f.}}  \simeq 0.1$ for total values):
\begin{multline}
    \frac{I_{\mathrm{magn/total}}^{\mathrm{gl, fit}}(T)}{T^4} = {} \\
    {} = \frac{I_{\mathrm{magn/total}}^{\mathrm{gl}, \infty}}{T^4} \left[1 + c_1 \frac{T_c}{T - T_c} + c_2 \left(\frac{T_c}{T - T_c} \right)^2 \right]\,, 
    \label{eq_magn_fit}
\end{multline}
\vspace{-1em}
\begin{align}
    I_{\mathrm{magn}}^{\mathrm{gl}, \infty}/T^4 & = 2.472(13),
    \quad c_1 = -0.999(5),
    \quad c_2 = 0.114(3),
    \nonumber \\
    I_{\mathrm{total}}^{\mathrm{gl}, \infty}/T^4 & = 3.74(19),
    \quad c_1 = -0.678(56),
    \quad c_2 = 0.079(24).
    \nonumber
\end{align}
These fits are shown by the solid lines in Fig.~\ref{fig_I}. 

Asymptotic results normalized by degrees of freedom do not coincide with those of the hot photonic gas. The reason is in the non-Abelian nature of magnetic degrees of freedom of gluons, which receive non-perturbative contributions even in the $T \to \infty$ limit~\cite{Karsch:2001cy}. Still, the hierarchy of these parts is the same: in the high-temperature regime, the purely magnetic degrees of freedom provide a dominant contribution to the total momentum. Still, just above the deconfinement transition at $T = T_c$, up to the supervortical critical temperature $T_s \simeq 1.5 T_c$, the negative contribution of the chromomagnetic gluons outweighs the positive contribution coming from the fluctuation of the angular momentum, and the plasma appears to carry a negative moment of inertia.

%
%
\section{Negative Barnett effect in gluon matter}
\label{ref_Barnett}

\subsection{Spin and orbital decomposition of the total angular momentum}

A negative moment of inertia implies that the total angular momentum ${\bs J} = {\bs L} + {\bs S}$, which (quark-)gluon plasma possesses, is pointed out in the opposite direction to the angular velocity ${\bs \Omega}$ with which the (quark-)gluon plasma rotates. How can this situation be realized in Nature? We put forward here the suggestion that the key player here is the gluon spin ${\bs S}$, which accumulates too much of total angular momentum ${\bs J}$, forcing, at the same time, the gluons themselves to rotate orbitally in the opposite direction with respect to the total angular momentum in order to compensate the over-excess of the angular momentum accumulated in the gluon spins ${\bs S}$. 

We call this effect the ``negative Barnett effect'' (NBE), which states that the spin polarization of gluons is opposite to the angular velocity of the system itself. The usual Barnett effect is the manifestation of the spin-orbital-coupling, which implies that the spin of the particles, experiencing a rotational collective motion, are aligned with the local angular velocity of the system~\cite{Barnett1915}.\footnote{The effect of the spin-orbital coupling caused by rotation should be contrasted to a similar effect that exists in {\it static} (nonrotating) bound systems. In a hydrogen atom, for example, the spin and orbital angular momentum of an electron tend to be anti-parallel, resulting in a lower energy of the ${}^2P_{1/2}$ state compared to the ${}^2P_{3/2}$ state. It is a fine-structure effect~\cite{Landau_QM_1976} caused by the polarization of an electron magnetic moment by a magnetic field arising in a (momentary) rest frame of the electron.} For ordinary uncharged substances such as certain metals~\cite{Barnett1917} or even water~\cite{Arabgol2019}, the spin polarization can experimentally be measured via rotation-produced magnetization of the sample.

The NBE, on the contrary, implies that the spin-orbital-coupling has a reversed sign so that the orientation of the spins~${\bs S}$ and the angular velocity ${\bs \Omega}$ is opposite to each other
within the supervortical window of temperatures $T_c \lesssim T < T_s$.
Namely, we suggest
that in a rotating gluon system: 
\begin{itemize}
    \item[(i)] a sizable fraction of the total angular momentum ${\bs J} = {\bs L} + {\bs S}$ is accumulated in the spin of gluons $\bs S$; 
    \item[(ii)] the spin polarization ${\bs S}$ is parallel to the total angular momentum ${\bs J}$ and anti-parallel to the orbital angular momentum of plasma $\bs L$. 
\end{itemize}

\begin{figure}[th]
    \includegraphics[width = 0.9\linewidth]{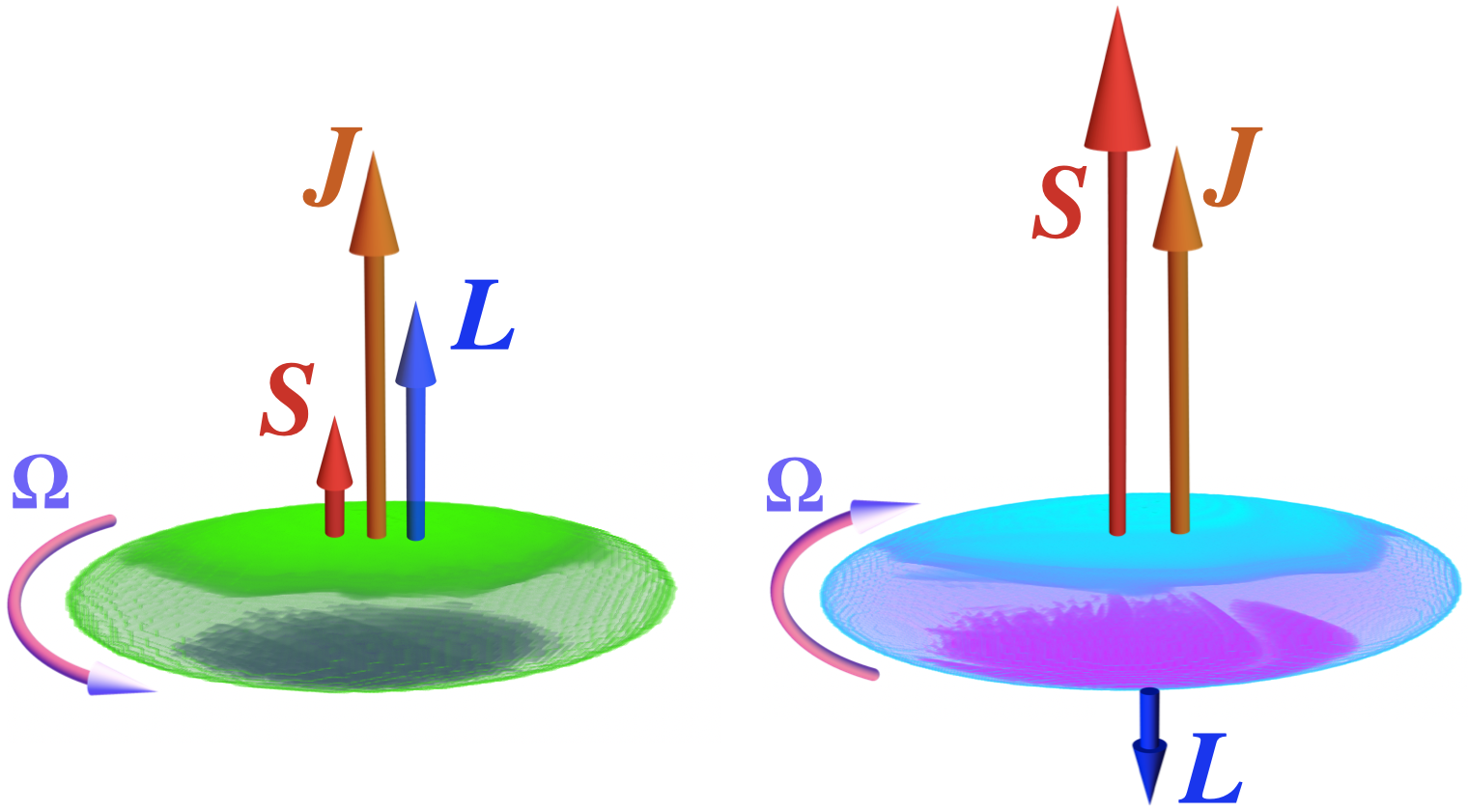}
    \caption{(left) The usual (positive) Barnett effect: the system possesses the positive angular momentum $\bs J$, which is redistributed between the positive orbital momentum $\bs L$, corresponding to the counterclockwise rotation and the positive spin polarization $\bs S$ generated by rotation. (right) The negative Barnett effect: the system possesses the positive angular momentum ${\bs J} = J {\bs n}$ (with ${\bs n}^2 = 1$), which is redistributed between a large positive spin polarization ${\bs S} = S {\bs n}$ (with $S > J > 0$) and an orbital angular momentum ${\bs L} = L {\bs n} = I_L {\bs \Omega}$ that takes a negative value $L < 0$ due to conservation of total angular momentum, $J = S + L$. Since the moment of inertia stored in orbital motion is positive, $I_L >0$, the system rotates, as a whole, clockwise: ${\bs \Omega} = \Omega {\bs n}$ with $\Omega < 0$. The negative Barnett effect implies that the angular velocity $\bs \Omega$ and spin polarization $\bs S$ are anti-parallel to each other.}
    \label{fig_NBE}
\end{figure}

Yet, in other words, projecting all angular momenta onto a single axis~$\bs n$ and assuming without loss of generality $J>0$, the gluonic spins accumulate a too large amount of angular momentum $S > J > 0$, so that this contribution needs to be compensated by a negative orbital motion, $L = J - S < 0$. Therefore, plasma acquires the angular momentum in one direction ($J>0$) and rotates in the opposite direction ($L < 0$), as illustrated in Fig.~\ref{fig_NBE}.

%
%

Notice that it is the orbital momentum that encodes the rotation of plasma as a whole, while the spin polarization is an intrinsic characteristic of the system that does not necessarily manifest itself in rotation. A relevant case is represented by a ferromagnet in the broken phase, which spontaneously generates a magnetic field due to substantial spin polarization, ${\bs S} \neq 0$, and, at the same time, does not rotate, ${\bs L} \equiv 0$~\cite{Landau_Statistics_1976}. Another example is given by superfluid ${}^3$He in a metastable but eternally long-lived $p$-wave state where the angular momentum, accumulated vortices, is opposite to the global rotation of the fluid~\cite{Autti2020}.

Given the suspected negativity of the spin-orbital coupling, one may ask ourselves how the moment of inertia is redistributed between spin and orbital degrees of freedom.
Following the analogy with the Barnett effect~\cite{Barnett1915}, we assert that the rotation with angular velocity ${\bs \Omega}$ induces the spin polarization of gluons, ${\bs S} = I_S {\bs \Omega}$.

Writing similarly for the orbital part, ${\bs L} = I_L {\bs \Omega}$, one arrives at Eq.~\eq{eq_L_Omega} where the total moment of inertia is the sum of the spin and orbital contributions: $I = I_S + I_L$. Since the energy density is a positive quantity, the mechanical orbital rotation can only lead to a positive value of the orbital moment of inertia, $I_L >0$. Therefore, the negative moment of inertia can only be achieved if the rotation polarizes spin negatively, $I_S < - I_L < 0$, against the rotation direction, so that $I = I_S + I_L < 0$.

%
%
\subsection{Implications for quark-gluon plasma}

So far, we have discussed the gluonic component of the rotating plasma. The magnetic condensate term will also exist in the presence of a rotating fermionic component so that our conclusions equally apply to the quark-gluon plasma. The coupling of quarks to rotation results in a linear term in the free energy, which implies that fermions themselves cannot generate the negative Barnett effect because their contribution to the total moment of inertia is associated with the single, purely mechanical term:
\begin{align}
    I^{{\mathrm{(q+gl)}}}_{\text{mech}} = \frac{1}{T} \aavr{ {\bs n} \cdot ({\bs J}_{\mathrm{q}} + {\bs J}_{\mathrm{gl}})^2}_T\,.
\label{eq_I_mech_q}
\end{align}
The angular momentum of quarks ${\bs J}_{\mathrm{q}}$ is given by Eq.~\eq{eq_angular_momentum} with the substitution $M^{ij}_{\mathrm{gl}} \to M_q^{ij} = i {\bar\psi} \gamma^0 (\gamma^i D^j - \gamma^j D^i) \psi$. The covariant derivative $D_\mu = \partial_\mu + i g A_\mu$ incorporates the coupling of the quark spinors $\psi$ to the gluon field~$A^\mu$.

The mechanical momentum, which includes the contribution of quarks~\eq{eq_I_mech_q}, is expected to give a positive addition to the total moment of inertia, thus diminishing the magnitude of its negative value in the supervortical range of temperatures. However, the evaporation of the gluon condensate also occurs in the presence of quarks, implying that the negative moment of inertia and the associated negative Barnett effect will also be realized in the quark-gluon plasma in relativistic heavy-ion collisions.

%
%
\section{Discussion and Conclusions}

It was recently found that the gluon plasma in the range of temperatures $T \simeq (1\dots 1.5) T_c$ possesses a negative moment of inertia, which may be interpreted as thermodynamic instability to the global rotation~\cite{Braguta:2023kwl, Braguta:2023yjn}. The result was obtained using two independent approaches in static~\cite{Braguta:2023kwl} and rotating~\cite{Braguta:2023yjn} plasmas. In this paper, we argued, using thermodynamic arguments and -- in relation to the gluon plasma -- the lattice data, that one can have a negative moment of inertia for an object with an everywhere-positive energy density.

Mathematically, the expression for the moment of inertia of a gluon system~\eq{eq_I} contains not only the expected mechanical term corresponding to the total angular momenta but also a local term that incorporates the chromomagnetic condensate. This additional term emerges due to the vector nature of gluons: the latter property leads to the non-linearity of the gluonic free energy, which contains terms both linear and quadratic in the angular frequency $\Omega$. The linear term produces the standard mechanical contribution~\eq{eq_I_mech} while the quadratic term leads to the appearance of the novel contribution, which incorporates the chromomagnetic condensate~\eq{eq_I_magn}. Our Monte Carlo simulation clearly shows that it is the thermal evaporation of the chromomagnetic condensate that leads to the negative moment of inertia of rotating plasma.  

We stress that the only reason for the emergence of a negative moment of inertia of gluons (Fig.~\ref{fig_I}) is the evaporation of the chromomagnetic condensate. Undoubtedly, the standard mechanical term~\eq{eq_I_mech} is a positively defined quantity. Therefore, the negative value of the condensate can only be a result of the negativity of the other term~\eq{eq_I_magn}, which is proportional to the thermal contribution to the chromomagnetic condensate. Our numerical data, in agreement with earlier studies in Ref.~\cite{Boyd:1996bx}, indeed indicates that the thermal chromomagnetic condensate is negative above the deconfinement phase transition (Fig.~\ref{fig_condensates}). The negative magnetic contribution~\eq{eq_I_magn} has a bigger amplitude than the (positive) mechanical part~\eq{eq_I_mech} in a specific range of temperature above the deconfinement phase transition, leading to the negative total value of the moment of inertia of spinning gluonic matter.

We could expect that a similar phenomenon can also appear in other field systems that satisfy two necessary requirements: 
\begin{itemize}
    \item[(i)] At zero temperature, the vacuum of the theory possesses a vacuum condensate, which evaporates as the temperature rises;
    \item[(ii)] The action of the theory possesses a quadratic coupling between the angular frequency $\Omega$ and the fields corresponding to such a condensate.
\end{itemize}
The second requirement leads to the emergence of a local contribution from the mentioned condensate to the moment of inertia. The first requirement assures that this term takes a negative value. In the case of Yang-Mills theory and QCD, such a term is the magnetic contribution of gluons~\eq{eq_I_magn} given by the chromomagnetic condensate. The quadratic terms in $\Omega$ appear to couple to the magnetic condensate in the Yang-Mills action.

Finally, let us consider fermions in rotation. Indeed, while the fermions do possess a chiral condensate that melts at high temperatures (the first criterion is satisfied), the coupling between rotation and fermionic bilinear is always a linear function of the angular frequency (see, for example, Refs.~\cite{Ambrus:2014uqa, Chen:2015hfc, Jiang:2016wvv, Chernodub:2016kxh}), which implies that the second criterion is not satisfied. Thus, the moment of inertia of pure fermionic matter is a positive quantity~\cite{Chernodub:2016kxh}.

Physically, we conjecture that the emergence of the Negative Barnett Effect appears due to too strong polarization of the gluon spins $\boldsymbol{S}$ in the direction of the total angular momentum $\boldsymbol{J}$ as visually illustrated in Fig.~\ref{fig_NBE}. We conclude that the gluon plasma mechanically rotates in the opposite direction to the angular momentum, implying the unusual relation ${\boldsymbol \Omega} \cdot {\boldsymbol J} < 0$. We coin this unexpected phenomenon as the Negative Barnett Effect.

%
%
\section{Acknowledgements}

This work has been carried out using computing resources of the Federal collective usage center Complex for Simulation and Data Processing for Mega-science Facilities at NRC ``Kurchatov Institute'', http://ckp.nrcki.ru/ and the Supercomputer ``Govorun'' of Joint Institute for Nuclear Research. 
The work of VVB, IEK, AAR, and DAS, which consisted in the lattice calculation of the observables used in the paper,  was supported by the Russian Science Foundation (project no. 23-12-00072). 
The work of MNC has been supported by the French National Agency for Research (ANR) within the project PROCURPHY ANR-23-CE30-0051-02. MNC is thankful to the members of Nordita for their kind hospitality. Nordita is supported in part by Nordforsk. The authors are grateful to Victor Ambru\cb{s} and Andrey Kotov for useful discussions.

\bibliography{moment}

\end{document}